\documentclass[12pt]{article}
 \usepackage{graphicx}
 \usepackage{amssymb,amsmath}
\begin{document}

\title{Neutralization of excited antiprotonic helium ion in collisions 
with helium atoms}
\author{G.Ya. Korenman and S.N. Yudin \\
\emph{\small Skobeltsyn Institute of Nuclear Physics,
 Lomonosov Moscow State University,}\\ \emph{\small Moscow 119991,  
Russia}\\
{\small E-mail: korenman@nucl-th.sinp.msu.ru}}

\date{}
\maketitle

\begin{abstract}
We consider a neutralization of excited antiprotonic helium ion accompanied 
by antiproton  transitions to lower states in collisions with helium atoms 
$(\bar{p}\mathrm{He}^{2+})_{nl} + \mathrm{He} \rightarrow \left[
(\bar{p}\mathrm{He}^{2+})_{n_f l_f} e\right]_{1s} + \mathrm{He}^+$ in 
low-temperature medium. Interactions in the input and output channels are taken 
similar to the potentials of $\mathrm{H}^+ - \mathrm{He}$ ($X^1\Sigma$)
and  $\mathrm{H} - \mathrm{He}^+$ ($A^1\Sigma$) systems, 
respectively. The final term is shifted down by the antiproton transition 
energy. Therefore initial and final terms can intersect for the suitable 
antiproton quantum numbers leading to an appreciable cross sections of the  
transitions. We have found that this process can provide an additional 
increase of effective annihilation rates for highly excited antiprotonic 
helium ion even in the low-density and temperature target corresponding to 
known experimental conditions. 

\end{abstract}

\section{Introduction} \label{sec1}
The experimental discovery and following study  of the metastable antiprotonic
states in helium have provided a large amount of data on various 
properties of this system,  especially  due to ASACUSA collaboration at 
Antiproton Decelerator in CERN
(see reviews \cite{ref1,ref2,ref3} and references therein).  High 
precision measurements of the laser-induced transitions from the 
metastable states and of the microwave-induced transitions  between 
hyperfine structure sublevels give an unique information on the 
fundamental properties of antiproton.
On the other hand, many other data, such as life times and populations 
of metastable states, shift and broadening of spectral lines \textit{etc.},
are connected with the mechanisms of formation and following 
interaction of antiprotonic helium with surrounding ordinary He atoms 
and, in special targets, with impurity atoms and molecules. 

Among many experimental results obtained by ASACUSA collaboration, there is one
very specific: the first observation of cold long-lived antiprotonic helium
ions $(\bar{p}\mathrm{He}^{2+})$ in the states with definite quantum numbers
($n=28\, - \,32, \,l=n-1$) \cite{ref4}. It was observed that these states have
lifetimes $\tau\sim 100$ ns against annihilation, and rates of annihilation
$\lambda=\tau^{-1}$ are increasing roughly as a linear function of target density at
$\rho<(5\, - \,10)\cdot 10^{17}$ cm$^{-3}$ and temperature $T\simeq 10$ K. The
antiproton annihilation occurs in the states with small angular momentum, 
therefore the observed lifetime has to be attributed to the time of
transitions from the initial  ($n,\,l$)-state to the $s,\,p,\,d$-states.
Radiative transitions from the circular orbits ($l=n-1$) at large $n$ are
strongly suppressed, therefore a main contribution to the effective rate of
annihilation comes from collisional transitions. In this case the observed
values of $\lambda$ (at the given $\rho$ and $T$) require to have an effective
cross section for the initial states $\sigma_q\sim (4\, - \,10)\cdot 10^{-15}$
cm$^2$. This value is an order of magnitude larger than the theoretical
cross sections for collisional Stark transitions obtained in previous 
calculations \cite{ref5, ref6, ref7} related to higher kinetic energy ($\sim$ 1 eV)
corresponding to $T\sim 10^4$ K. In addition to huge quenching cross 
sections, the experiment revealed an unexpected $n$-dependence and isotope 
effect ($A=3,\,4$) for effective annihilation rates. These results also 
can not be explained by earlier considerations of collisional processes of 
$(\bar{p}\mathrm{He}^{2+})$.
 
In the papers \cite{ref8, ref9} collisions of cold 
$(\bar{p}\mathrm{He}^{2+})_{nl}$ ion with surrounding helium atoms at $T\sim 
10$ K were considered in the framework of close coupling  approach with 
account for the states with all angular momenta $l$ from $n-1$ to 0 at the 
fixed $n$. A model interaction between two colliding subsystems was taken as 
an adiabatic potential $V_0(R)$ of the interaction between He atom and unit 
charge of the ion. Transitions between the states of $(\bar{p}\mathrm{He}^{2+}
)_{nl}$ ion are produced by the interactions of electric dipole and 
quadrupole momenta of the ion with the electric field $\mathbf{E}(\mathbf{R})
\sim -\nabla_{\mathbf{R}}V_0(R)$ of the polarized He atom.  This model gives 
cross sections of Stark transitions from circular orbits with $n\sim 30$ at $E
=10$ K that are of order or greater than effective quenching cross section $
\sigma_q$. Stark transition rates averaged over thermal motion correlate 
similarly with the observed effective annihilation rates. Thus, the model 
allows to understand 
qualitatively the observed data. Moreover, it gives also some isotope effect 
with correct sign and an increasing of the rates with $n$. However, if to 
take into account a whole cascade of Stark and radiative transitions from the 
initial circular orbit up to annihilation from $s$ and $p$-states, it turns 
out that the values of obtained Stark cross sections are not large enough. An 
improvement of the calculated Stark transition rates, apparently, can be 
obtained by using the potential energy surface for the two-electron system $
\bar{p}\mathrm{He}^{2+}\,-\,\mathrm{He}$ \cite{ref10}. 

On other hand, for the 
further understanding and better quantitative description of the observed 
data it is interesting to search possible additional mechanisms of 
collisional quenching of antiprotonic helium ion. One class of such mechanisms 
can be antiproton transitions accompanied by energy transfer to the electron 
of He atom, such as
\begin{align}
(\bar{p}\mathrm{He}^{2+})_{nl} + \mathrm{He} &\rightarrow 
(\bar{p}\mathrm{He}^{2+})_{n_f l_f} +e  + \mathrm{He}^+  ,
 \label{eq1} \\
(\bar{p}\mathrm{He}^{2+})_{nl} + \mathrm{He} &\rightarrow 
\left[(\bar{p}\mathrm{He}^{2+})_{n_f l_f} e\right]_{1s} + \mathrm{He}^+ .
 \label{eq2} 
\end{align} 
First process usually referred as external Auger transition. Second process 
is a neutralization of antiprotonic helium ion, or electron transfer from the 
neutral He atom to the ion. Two processes seem very similar, differing only 
by electron final states. 
However at low-temperature collisions the external Auger process is strongly 
suppressed by Coulomb repulsion of two positive  heavy ions in the final 
state, whereas in the second process this suppression is absent due to 
neutralization of the ion. In addition, Auger process suppressed also due to 
higher multipolarity $\Delta l= l - l_f$ of antiproton transitions that 
follows from the energy conservation conditions. Kinetic energy of colliding 
subsystems in the low-temperature target can be neglected, therefore the 
energy-allowed Auger transitions have to satisfy the condition
 \begin{equation} \label{eq3}
- \frac{\mu Z^2}{2 n^2} - I_0  \geq  
- \frac{\mu Z^2}{2(n-\Delta n)^2} + \varepsilon_e ,
\end{equation}
where $\mu=m_{\bar{p}}M_a/(m_{\bar{p}}+M_a)$ is the reduced mass of 
antiproton-nucleus system, $Z=2$, $I_0 \simeq 0.9034$ a.u. is the ionization 
energy of He atom, $\Delta n = n - n_f$,  $\varepsilon_e \geq 0$ 
is the electron energy in the final state. Here and further in this work we 
use atomic units ($\hbar=e=m_e=1$) unless otherwise specified. 
At $n=30$ the condition \eqref{eq3} requires for the allowed Stark 
transitions $\Delta n \geq 4$.  For the circular orbits  $\Delta l=\Delta n$, 
therefore a minimum multipolarity of Auger transition from 
the (30,\,29) state is also equal 4.

In order to obtain minimum multipolarity of antiprotonic transition during 
neutralization, we need to know the energy $E_{n_f l_f}$ of the neutral 
antiprotonic helium atom in final state with 
$l_f\leq n_f-1,\, n_f=n-\Delta n\lesssim 30$. Energy levels of this system 
were calculated with high precision \cite{ref11, ref12} for many states with $
n>30$ related to laser-induced transitions. However, final states in 
the problem under consideration will have smaller quantum numbers $n_f$.
Therefore, for rough estimations we can begin from the 
simple model \cite{ref13} of hydrogen-like wave functions with effective 
charges $Z_{\bar{p}}$ for antiproton and $Z_e$ for electron. The mean radii of 
electronic and (circular) antiprotonic orbits are $\bar{r}_e=3/(2Z_e)$ and 
$\bar{r}_{\bar{p}}=n(n+1/2)/(\mu Z_{\bar{p}})$, respectively. Variational 
calculations for circular orbits within this model \cite{ref13} show that
$Z_e \lesssim 1.15$, $0.95<Z_{\bar{p}}<2$ at $n\leq 30$, hence 
$\bar{r}_e\simeq 1.5$, $\bar{r}_{\bar{p}} \leq 0.3$. 
Therefore we can simplify the model considering a system at $n \lesssim 30$ 
in a zero approximation as antiproton in Coulomb field of nucleus charge +2 and 
electron in the field of charge +1 placed into the center of mass of 
antiproton-nucleus system.
The residual interaction
\begin{equation} \label{eq4}
V_{res} = - 2/|\mathbf{r}_e +\nu \mathbf{r}| + 1/|\mathbf{r}_e -\lambda 
\mathbf{r}| + 1/r_e ,
\end{equation}
will be taken into account in the first-order perturbation theory. Here 
$\lambda=M_a/(M_a +m_{\bar{p}})$, $\nu=m_{\bar{p}}/(M_a +m_{\bar{p}})$, $
\mathbf{r}$ is the antiproton coordinates counted from the nucleus and $
\mathbf{r}_e$ is electron coordinates counted from antiproton-nucleus c.m. In 
this approximation the energy of the neutral
antiprotonic atom  is estimated as
\begin{equation} \label{eq5}
E_{n_f l_f}= - \frac{\mu Z^2}{2n_f^2} -\epsilon_{1s} -\Delta \epsilon_{1s}(n_
f l_f),
\end{equation}
where $\epsilon_{1s}=1/2$ is the bounding energy of electron in 1s-state, and 
$\Delta \epsilon_{1s}(n_f l_f)$ is a correction to 
the bounding energy due to a residual interaction,
\begin{equation} \label{eq6} 
\Delta \epsilon_{1s}(nl)= -\big{\langle} 1s,n l\big{|} V_{res} 
\big{|}1s,n l \big{\rangle}\simeq  \frac{\lambda^2- 2 \nu^2}{3(\mu Z)^2}
n^2[5n^2 +1 -3l(l+1)]
\end{equation} 
An approximate expression in the right side of \eqref{eq6} is obtained at 
$\bar{r}_e^2 \gg \bar{r}_{\bar{p}}^2$.

Now the condition for energy-allowed transitions in the case of the 
neutralization process \eqref{eq2}  can be written as
 \begin{equation} \label{eq7}
-\frac{\mu Z^2}{2n^2} - I_0  \geq 
-\frac{\mu Z^2}{2(n-\Delta n)^2} - (\epsilon_{1s} + \Delta \epsilon_{1s})
\end{equation}
It differs from Eq. \eqref{eq3} by the negative electron energy 
($-\epsilon_{1s} -\Delta \epsilon_{1s}$) in the final state, allowing the 
transitions with $\Delta n \geq 2$ at $n_f\lesssim 30$. Respectively, 
minimum multipolarity of transitions is $\Delta l=2$  for initial circular orbit
and $\Delta l=1$ for nearly-circular ($l\leq n-2$) orbits, contrary to 
$\Delta l = 4$ for external Auger transitions. (For the states with $l\leq n-3$,
in principle, monopole transitions with $\Delta l =0$ can be also allowed, but,
as usually, they are strongly suppressed.) 

In this paper we consider a neutralization \eqref{eq2} of excited 
antiprotonic helium ion accompanied by antiproton  transitions to lower 
states in collisions with helium atoms in low-temperature medium. 
Interactions in the input and output channels are 
taken similar to the potentials of $\mathrm{H}^+ - \mathrm{He}$ ($X^1\Sigma$)
and  $\mathrm{H} - \mathrm{He}^+$ ($A^1\Sigma$) systems, respectively. The 
final term is shifted down due to a change of the antiproton energy. Therefore 
initial and final terms can intersect for the suitable antiproton quantum 
numbers leading to an appreciable cross sections of the  transitions. We have 
found that this process can provide an additional 
increase of effective annihilation rates for highly excited antiprotonic 
helium ion even in the low-density and temperature target corresponding to 
known experimental conditions. 

 The used approach is outlined in the next Section \ref{sec2}. Approximations 
for the cross sections and transition rates of the neutralization are considered 
in Sec. \ref{sec3}. Results of calculations are shown in Sec. \ref{sec4}.
A general conclusion is given in Sec. \ref{sec5}.

\section{Approach to description of the neutralization process} \label{sec2}

Let us consider an interaction $V_i(R)$ of the colliding subsystems in the 
input channel of the reaction \eqref{eq2}. At large and intermediate distances 
($R\gg n^2/(\mu Z) \simeq 0.3$) it can be approximated as interaction between 
heavy structureless unit charge +1 with He atom corresponding to $X^1\Sigma$ 
term of (p-He) system. Therefore we take a potential energy curve (PEC) 
$U_{X^1\Sigma}(R)=V_{\mathrm{H^+ He}}(R)$ as the potential in input channel
\begin{equation} \label{eq8}
 V_i(R) = V_{\mathrm{H^+ He}}(R) . 
\end{equation}
The data for this PEC obtained by \emph{ab initio} calculations were taken 
from the paper \cite{ref14} (see also \cite{ref15}). Energy of the term 
$X^1\Sigma$ counted from the ground state of He atom, therefore at very large 
distances this potential tends to zero as $-\alpha/(2R^4)$, 
where $\alpha$ is a polarizability of He atom.

After the electron transfer, an interaction between two subsystems in output 
channel at large $R$ looks like $(\mathrm{H} - \mathrm{He}^+)$ interaction. 
We take the corresponding potential energy curve as 
$U_{A^1\Sigma}(R)=V_{\mathrm{H He}^+}(R) + I_0 -\epsilon_{1s}$, 
where the potential $V_{\mathrm{H He}^+}(R)$ of ($\mathrm{H-He^+}$) interaction 
was taken from the paper \cite{ref16} 
\footnote{It should be noted that due to electron transfer the relative 
distances between two subsystems in input and output channels are, strictly 
speaking, differ by value of order $m_e/M_a$. It can be essential in the 
asymptotic behaviour of radial wave functions. However, we will see below 
that a probability of electron transfer in the used approximation is 
calculated at the region of interaction, therefore we don't need to consider these 
asymptotics.}.
The potential energy curve $U_{A^1\Sigma}(R)$ goes to $I_0 -\epsilon_{1s}$ at 
$R\rightarrow \infty$ and, of course, does not cross the potential curve  
$U_{X^1\Sigma}(R)$. However, the inner energy of two subsystem  
($\left[(\bar{p}\mathrm{He}^{2+})_{n_f l_f} e\right]_{1s}$ + $\mathrm{He}^+$)
 is shifted down due change of antiproton state and a correction $\Delta \epsilon_{1s}$ to electron bounding energy. Therefore we take the interaction potential in output 
channel as 
\begin{equation} \label{eq9}
V_f(R)= V_{\mathrm{H He}^+}(R) + I_0 -\epsilon_{1s} -  \Delta \epsilon_{1s}  
-\frac{\mu Z^2}{2} \left(\frac{1}{(n-\Delta n)^2} -\frac{1}{n^2}  \right )
\end{equation} 
Then, with a suitable choice of quantum numbers, curves $V_f(R)$ and $V_i(R)$ 
can intersect.

\begin{figure}[thb]
\includegraphics[width=\textwidth]{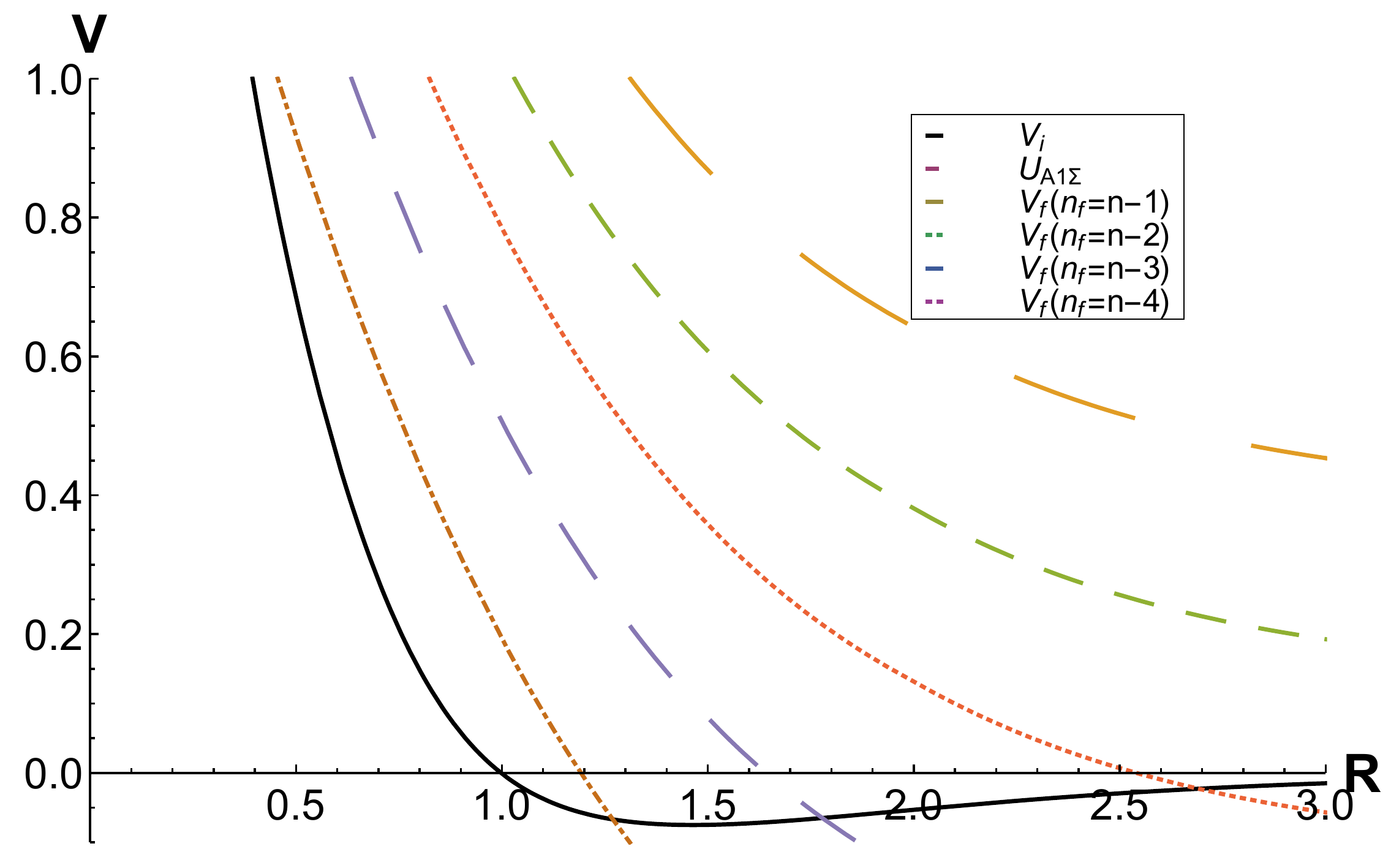} 
 \caption{Potentials of interaction between two subsystems in input and output 
channels of the reaction \eqref{eq2} for initial antiprotonic state $n=30$ and different final states} 
    \label{fig1}
\end{figure}

Typical results for the crossing curves are shown in Fig. \ref{fig1}. It is seen
from the figure that transitions with $\Delta n=1$ in thermal collisions  are
forbidden by the energy conservation. Potentials $V_f(R)$ at $\Delta n=2,\,3,\,4$
crosses $V_i(R)$ in the region $R>1$ corresponding to attraction in the input  
channel, whereas at $\Delta n= \geq 5$ possible crossing points (not shown in 
Fig. \ref{fig1}) fall into the repulsion region $R<1$, which is classically 
forbidden in the input channel at thermal energy. Therefore we can restrict our
consideration of the neutralization process with $\Delta n$ from 2 to 4. 
Moreover, higher $\Delta n$ requires higher multipolarity that can strongly suppress a probability of transition.  

\section{Approximations for cross-sections} \label{sec3}
 
It is well known that the main contribution to transitions during low-velocity collisions gives an area nearly to point of crossing between the 
therms of initial and final channels. Therefore, using Landau-Zener 
approximation \cite{ref17,ref18} we can estimate cross section of electron 
transfer \eqref{eq2} by equation
\begin{equation} \label{eq10}
\sigma_{x}(E) = 2\pi \int_0^{b_m (E)} P_{fi}(b,E) \left[1 - P_{fi}(b,E)\right]
b\,db ,
\end{equation} 
where $P_{fi}(b,E)$ is a probability of transition during single passing of 
the system through the crossing point. It can be written as
\begin{align} 
P_{fi}(b,E) &= \exp[-\zeta(b,E)], \label{eq11} \\
\zeta(b,E)  &=\frac{2\pi |V_{fi}(R_x)|^2}{v(R_x)|V_i '(R_x) - V_f '(R_x)|}
,    	\label{eq12}	
\end{align}
where $V_{fi}(R)$ is a matrix element of interaction that couples channels, 
$R_x$ is a crossing point of potentials $V_i(R)$ and $V_f(R)$, $v(R)=\sqrt{2\left[E-
V_i(R)-E b^2/R^2 \right]/M}$ is a local (classical) velocity of relative 
motion at the point $R$, $b$ is impact parameter, and E is energy of collision. 
Upper limit $b_m(E)$ of integration in Eq. \eqref{eq10} is a maximum impact parameter allowing the system to approach the crossing point that correspond to the condition on the most right classical turning point $R_t(E,b_m)=R_x$.

The process under consideration \eqref{eq2} involves electron transfer and 
simultaneous  antiproton jump to a lover orbit, therefore the transition 
operator has to contain both electronic and antiprotonic variables. It 
corresponds to non-diagonal part of Coulomb interaction between electrons and 
antiprotonic ion constituents, 
\begin{equation} \label{eq13}
V^{tr}(\mathbf{R}, \mathbf{r}, \mathbf{r}_1, \mathbf{r}_2) = \sum_{e=1,2}\Big(
\frac{1}{|\mathbf{R}-\lambda\mathbf{r}+\mathbf{r}_e|} -
\frac{2}{|\mathbf{R} +\nu\mathbf{r}+\mathbf{r}_e|} \Big).
\end{equation}
 In line with the adapted approximation for $V_i(R)$ and $V_f(R)$, 
initial and final wave functions of the system can be presented as
\begin{align}
|\,i\,\rangle & = \Phi_{nlm}(\mathbf{r})
	\Psi_{X^1\Sigma}(\mathbf{r}_1, \mathbf{r}_2;R), \label{eq14}  \\
|f\rangle & = \Phi_{n'l'm'}(\mathbf{r})
\Psi_{A^1\Sigma}(\mathbf{r}_1, \mathbf{r}_2;R), \label{eq15}
\end{align}
were $\Phi_{nlm}(\mathbf{r})$ and $\Psi_{^1\Sigma}(\mathbf{r}_1, \mathbf{r}_2;R)$
are wave functions of antiproton and of two electrons, the latter being
dependent on the distance between two subsystems. Indexes $X^1\Sigma$  
and $A^1\Sigma$ indicate that initial and final electronic wave functions 
correspond to ground state of $\mathrm{He} + \mathrm{H}^+$ system and to 
$\mathrm{He}^+ + \mathrm{H}$ system, respectively.
 
Detailed calculation of the transition matrix element is quite complicated by  
necessitate to know an extensive set of parameters for the initial and final
electronic wave functions at different values of $R$. However, we can estimate 
this matrix element taking in mind that electronic 
$A^1\Sigma \rightarrow X^1\Sigma$ transition dipole moment $D(R)$ was 
tabulated in Ref. \cite{ref16} as a function of $R$. For this aim we 
approximate non-diagonal part of the $V^{tr}$ by the interaction of electronic 
dipole with antiprotonic dipole and quadruple,
\begin{equation} \label{eq16}
\begin{split}
V^{tr}(\mathbf{R}, \mathbf{r}, \mathbf{r}_1, \mathbf{r}_2) &\simeq (\lambda+2
\nu)
\big[(\mathbf{d_e\, r})- 3 (\mathbf{d_e\, n})(\mathbf{r\, n})\big] \big /R^3\\
  & + 3(\lambda^2-2\nu^2) \big[(\mathbf{d_e\, n}) r^2 + 2(\mathbf{d_e\, r}) 
(\mathbf{r\, n}) - 5 (\mathbf{d_e\, n}) (\mathbf{r\, n})^2 \big]/(2R^4),\\
\end{split}
\end{equation}
where 
 $\mathbf{d_e}=\sum_e \mathbf{r_e}$ and $\mathbf{n}=\mathbf{R}/R$. 
Matrix element of the vector $\mathbf{d_e}$ is directed along $\mathbf{n}$
due to axial symmetry of the both $^1\Sigma$ terms, 
\begin{equation} \label{eq17}
\langle A^1\Sigma|\mathbf{d_e}| X^1\Sigma \rangle = D(R)\,\mathbf{n},
\end{equation} 
therefore, we can write
\begin{equation} \label{eq18}
\begin{split}
V_{fi}(\mathbf{R}) \simeq &- 2 D(R)(\lambda+2\nu) 
\langle l'm'|\cos\theta|lm \rangle I_1(nl,n'l') /R^3 \\
  & - 3 D(R)(\lambda^2-2\nu^2)
\langle l'm'|P_2(\cos\theta)|lm \rangle I_2(nl,n'l') /R^4 , \\
\end{split}
\end{equation}  
where 
\begin{equation} \label{eq19}
I_k(nl,n'l') = \int_0 ^\infty R_{nl}(r)R_{n'l'}(r) r^{k+2}\,dr ,  
\end{equation}
$R_{nl}(r)$ are hydrogen-like radial wave functions of antiproton,
$\cos\theta=(\mathbf{n\cdot r})/r$. 

Strictly speaking, in addition to the last term in Eq. \eqref{eq18}, one should 
also consider another term  with the same 
$R^{-4}$-dependence corresponding to interaction between antiproton dipole 
and electron quadruple momenta. The latter is unknown therefore we need to 
omit it keeping the quadruple term in  Eq. \eqref{eq18} in order to estimate a 
contribution from antiproton quadruple transitions accompanied electron dipole 
transitions.
 
We have seen above (see Fig. \ref{fig1}) that a neutralization of 
antiprotonic helium ion with $n\sim 30$ is allowed by energy in the case $
\min\Delta n =2$.  For initial circular orbit it means also $\min\Delta l=2$, 
but for $l\leq n-2$ allowed transitions have $\min\Delta l=1$.
As a first step, we consider in this paper neutralization cross sections only 
for transitions with $\Delta n=2$.

Angular matrix elements of $\cos\theta$ and $P_2(\cos\theta)$ in Eq. 
\eqref{eq18} can be calculated in terms of Clebsch-Gordan coefficients, and then
an averaged square of transition matrix element
\begin{equation}  \label{eq20}
\overline{|V_{fi}(R)|^2}\equiv \frac{1}{2l+1} \sum_{m\,m'}|V_{fi}(\mathbf{R})|^2
\end{equation}
is reduced to the following form
\begin{equation} \label{eq21}
\begin{split}
 \overline{|V_{fi}(R)|^2} &=\frac{4}{3} D^2(R) (\lambda+2\nu)^2 
\langle l\,0\,1\,0|l'\,0\rangle^2 I_1^2(nl,n'l')\big/R^6 \\
& + \frac{9}{5} D^2(R) (\lambda^2-2\nu^2) 
\langle l\,0\,2\,0|l'\,0\rangle^2 I_2^2(nl,n'l')\big/R^8 ,
\end{split}
\end{equation} 
where first and second terms correspond to dipole and quadruple antiproton 
transitions. Of course, an interference term is absent here due to parity 
selection rule.

\section{Results for neutralization rates} \label{sec4}

\begin{figure}[thb]  
\centering
    \includegraphics[width=0.75\textwidth]{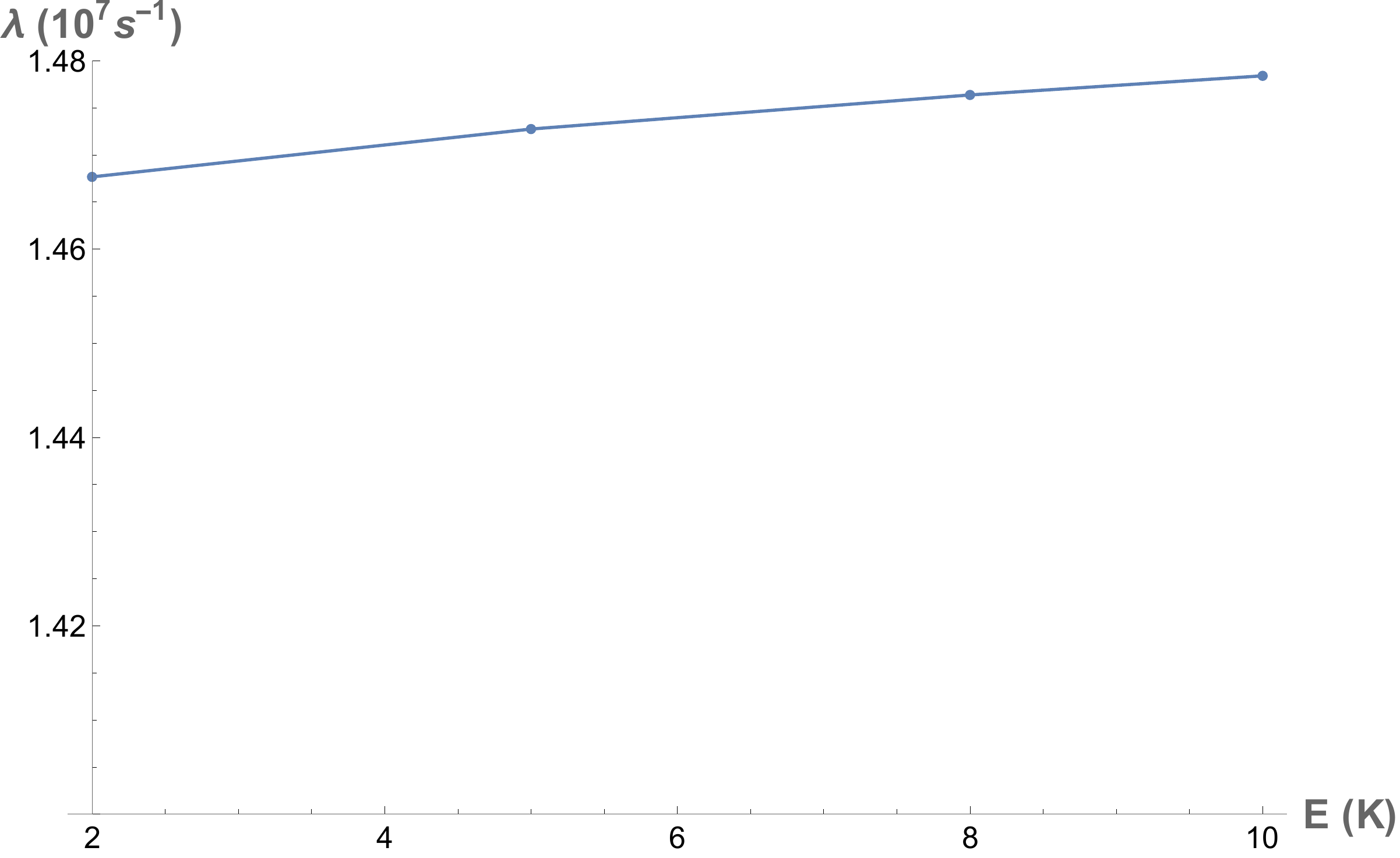}
\caption{Energy dependence of the neutralization rate
$\lambda(i\rightarrow f) =\rho\sigma(i\rightarrow f) v $
for antiproton dipole transition $(30,28\rightarrow 28,27)$ at the target 
density $\rho=5\times 10^{17}\,cm^{-3}$.}
    \label{fig2}
\end{figure}

Using the above-mentioned approximations we have calculated cross sections and
transition rates of the neutralization process \eqref{eq2} for initial
antiprotonic helium ion in the states with $n$ from 28 to 32 and different 
$l\leq n-1$ at energies nearly to the temperature of the target (T = 10 K) in 
the experiment \cite{ref4}. In order to compare the neutralization process with 
the radiative decay of antiprotonic helium ion, we show results for the 
transition rates
\begin{equation} \label{eq22}
 \lambda(i\rightarrow f) = \rho\sigma(i\rightarrow f) v
\end{equation}
Typical energy dependence of the transition rate is shown in Fig. \ref{fig2} for
antiproton dipole transition $(30,28\rightarrow 28,27)$ at the target density 
$\rho=5\times 10^{17}\,cm^{-3}$. It is seen that  this rate is very weakly depends 
on energy, therefore an averaging over thermal motion can be  replaced by the 
rate value at $E=T$. 

\begin{figure}[thb]
\centering
    \includegraphics[width=0.75\textwidth]{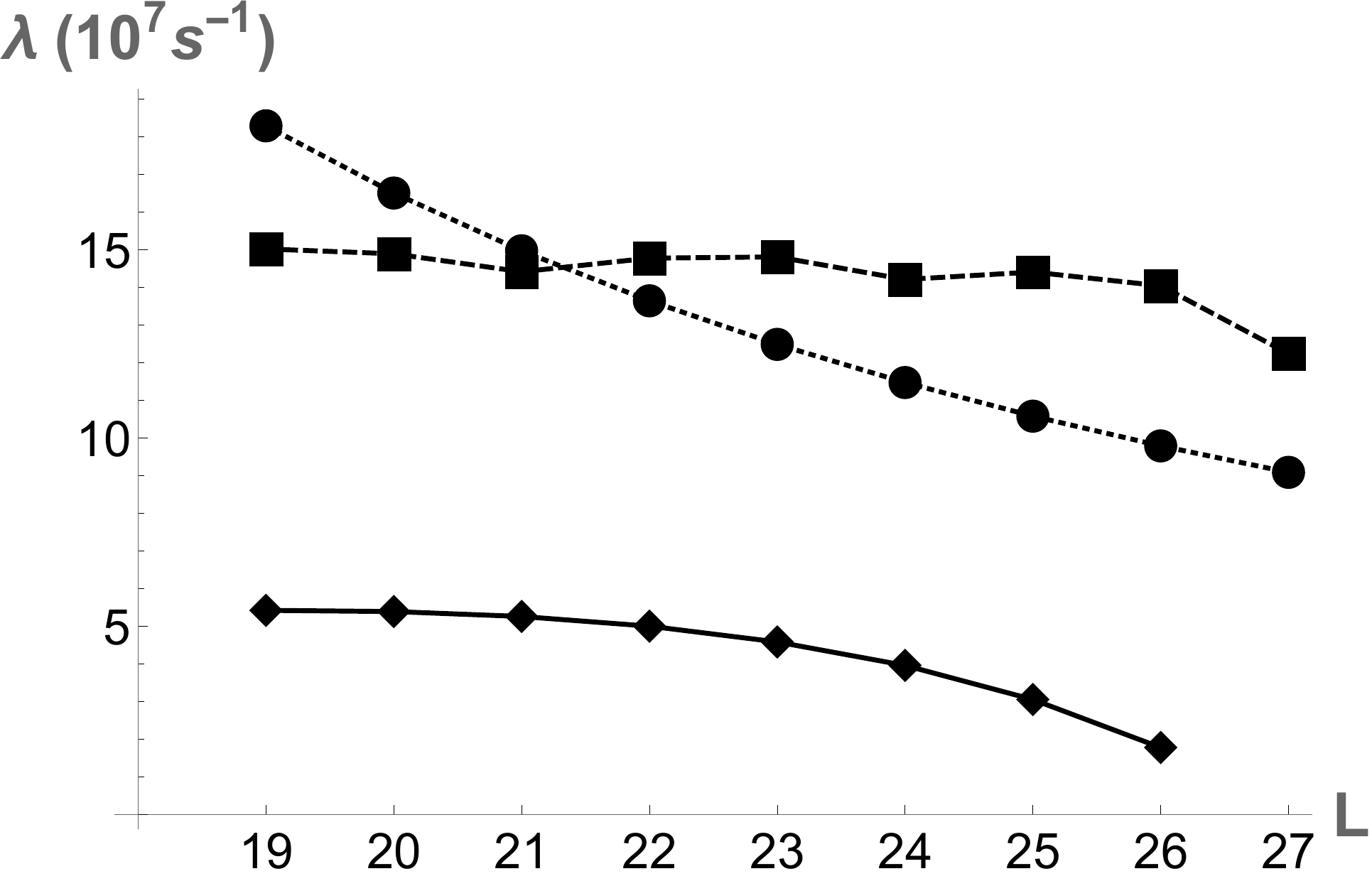}
		\caption{Rates of transitions from the states with $n=28$. 
		Legend: dashed curves (squares) are for Stark transitions, dotted curves 
		(solid discs) for radiative transitions, and solid curve for neutralization.}
		\label{fig3}
	\end{figure}

\begin{figure}[thb]
\centering
    \includegraphics[width=0.75\textwidth]{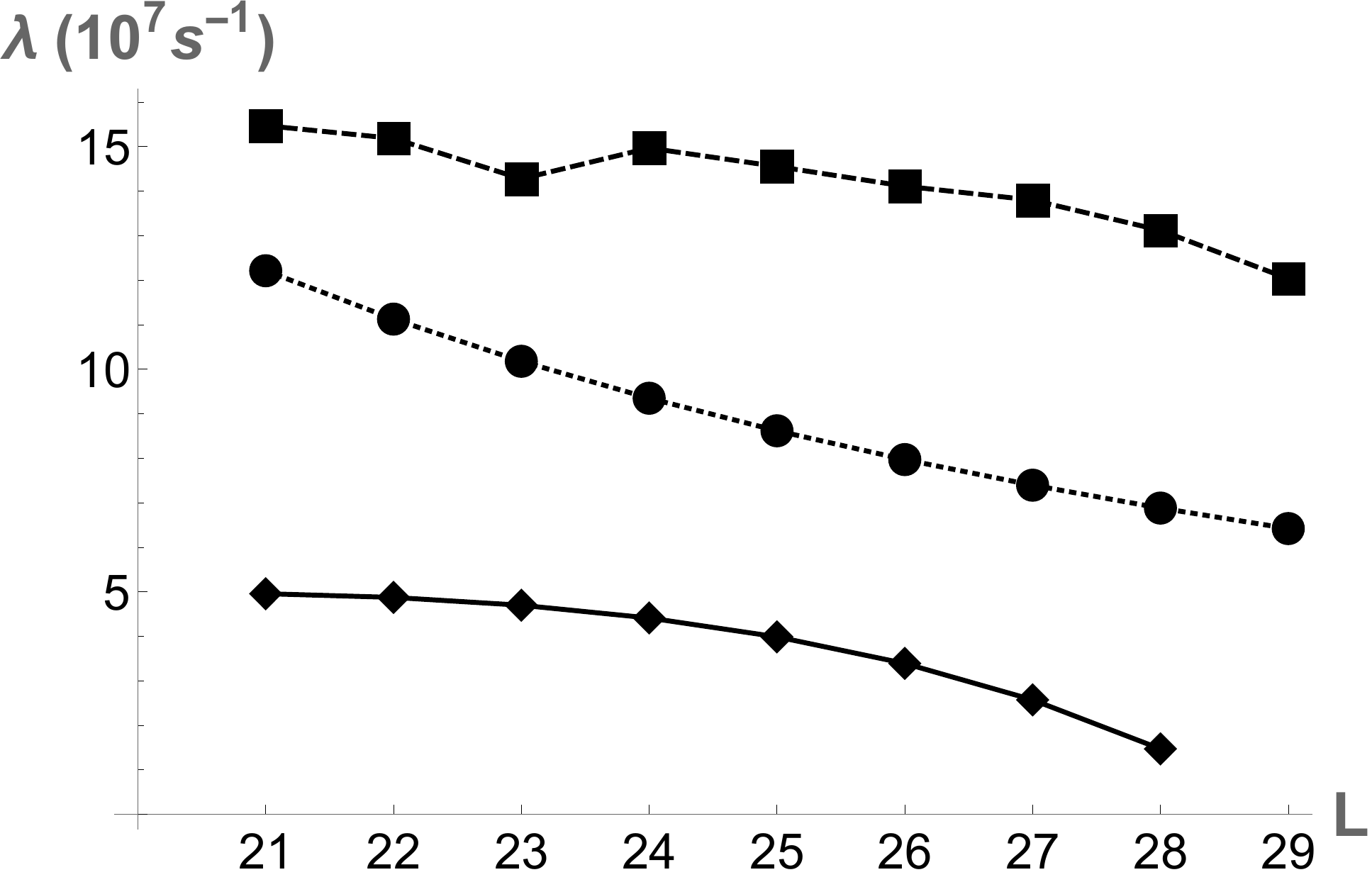}
\caption{Rates of transitions from the states with $n=30$ depending on initial 
angular momentum. Legend is the same as in Fig. \ref{fig3}.}
		\label{fig4}
	\end{figure}

\begin{figure}[thb]
\centering
    \includegraphics[width=0.75\textwidth]{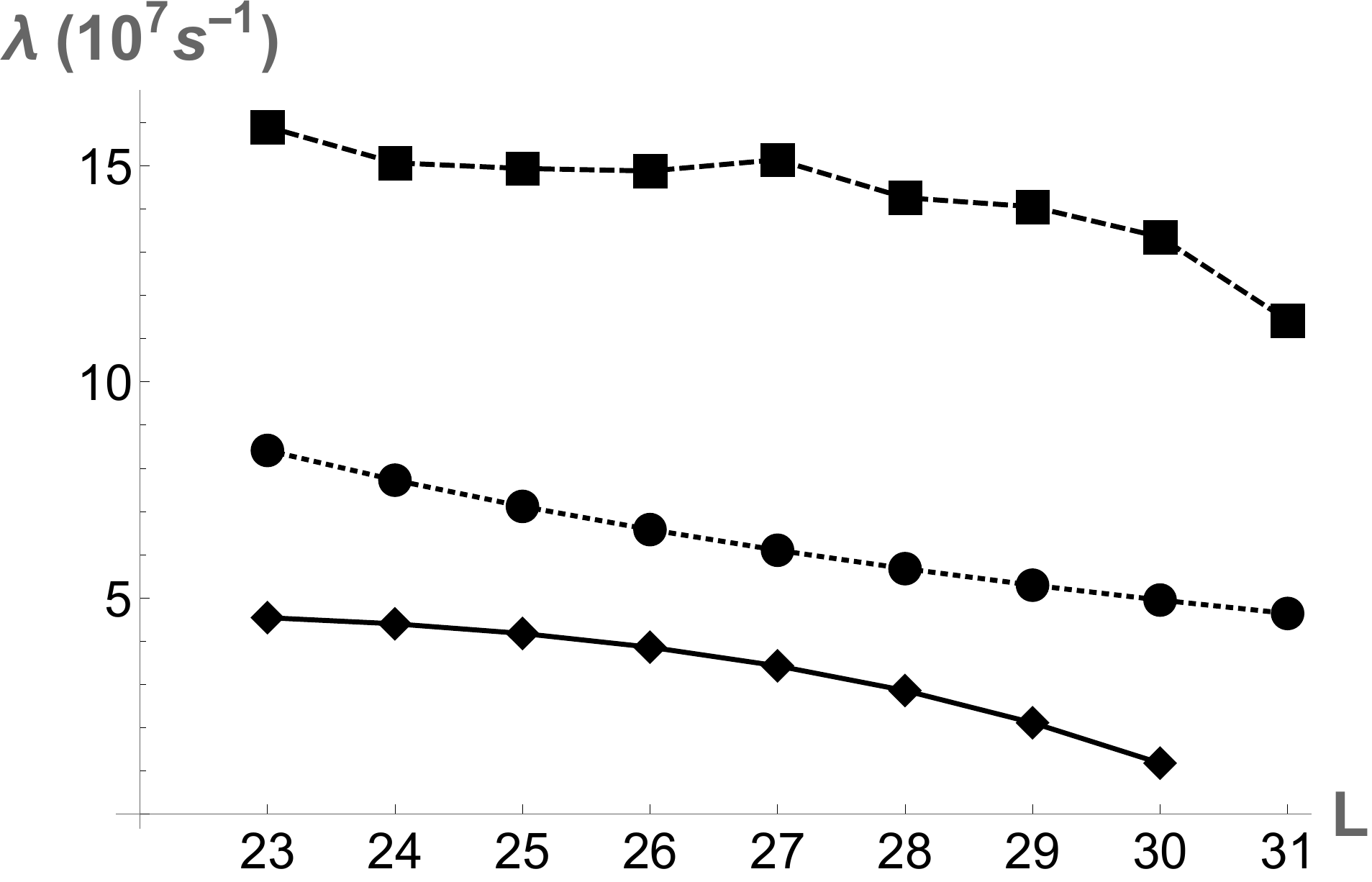}
	\caption{Rates of transitions from the states with $n=32$ depending on 
	initial angular momentum. Legend is the same as in Fig. \ref{fig3}.}
		\label{fig5}
	\end{figure}

Figs. \ref{fig3} - \ref{fig5}  show the neutralization rates for initial states
$n=28,\,30,\,32$ depending on initial angular momentum $l$ at $T=10\,K$ and the
same target density ($\rho=5\times 10^{17}\,cm^{-3}$). For a comparison, rates of 
radiative transitions and summary rates of Stark transitions \cite{ref8,ref10} 
from the state $(n,l)$ to all $(n,l'<l)$ are also given on the same figures.
The shown neutralization rates correspond to dipole antiprotonic transitions 
from non-circular orbits. As for circular orbits, the dipole transitions are 
forbidden by energy, whereas quadruple ones are suppressed at least by one 
order of value, therefore they are not shown in the scale of figures.

\begin{figure}[thb]
\centering
    \includegraphics[width=0.75\textwidth]{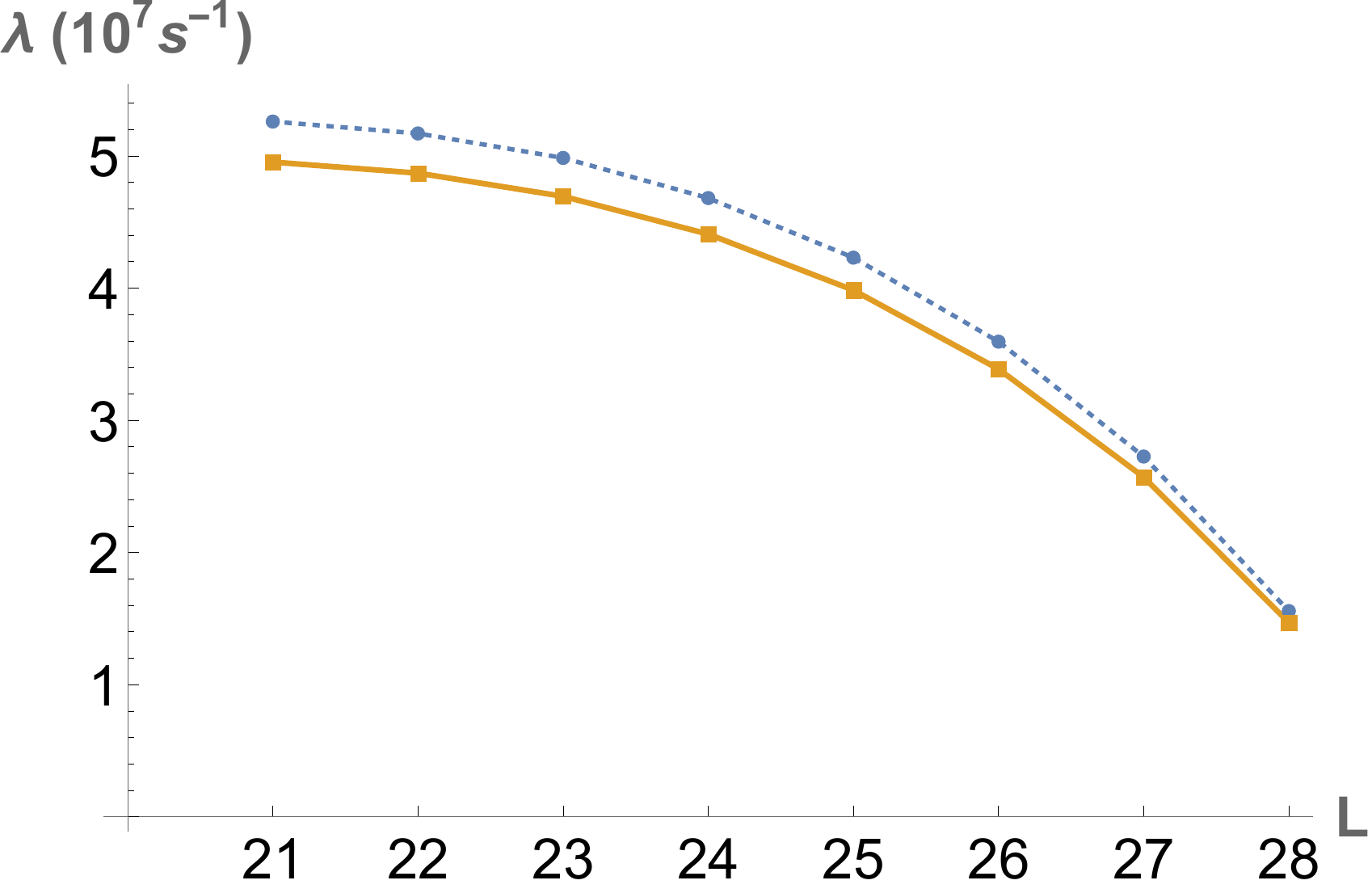}
	\caption{Neutralization rates for two antiprotonic ions 
	$(\bar{p}^3\mathrm{He}^+)$ (dashed line) and $(\bar{p}^4\mathrm{He}^+)$ 
	(solid line) depending on antiproton initial angular momentum at $n=30$, in 
	the target with $T=10\,K$, $\rho=5\times 10^{17}\,cm^{-3}$.}
		\label{fig6}
	\end{figure}

In addition to the $(\bar{p}^4\mathrm{He}^+)$ neutralization, we have considered 
also a similar process for $(\bar{p}^3\mathrm{He}^+)$. Fig. \ref{fig6} show a 
comparison of neutralisation rates for two isotopic targets with $T=10\,K$, 
$\rho=5\times 10^{17}\,cm^{-3}$ at $n=30$ depending on initial antiproton angular 
momentum.  A sign of the isotope effect coincides with the experimental 
data, i.e. rates of transitions for $^3$He are greater than for $^4$He, however
the value of isotopic effect is rather small, especially for 
nearly-circular states. It is slowly increasing to lower $l$ and approaches to 
10\% at $l=20$.

\section{Conclusion} \label{sec5}

We have considered process of antiprotonic helium ion neutralization accompanied
by antiproton  transitions to lower states in collisions with helium atoms 
$(\bar{p}\mathrm{He}^{2+})_{nl} + \mathrm{He} \rightarrow \left[
(\bar{p}\mathrm{He}^{2+})_{n_f l_f} e\right]_{1s} + \mathrm{He}^+$ in 
low-temperature medium. This process can be referred also as charge exchange 
between ion and atom. A simple model of the process is developed allowing for 
to estimate cross sections and transition rates. It is shown that the 
neutralization gives a remarkable contribution  to summary rates of antiproton 
going from the states with $n=28 - 32$ to lower states. Rates of this process reach up to 50\% of radiative transition rates. Isotopic effect in the process 
has the same sign as in experiment, however it's value is rather small.
As a general conclusion, it was shown that the neutralization process has to be taken into account together with the radiative transitions in the cascade 
calculations of effective annihilation rates for antiprotonic helium ion.

\end{document}